\documentclass[aps,prb,showpacs,twocolumn,balance]{revtex4-2}

\usepackage{newtxtext}

\usepackage{subfigure}
\usepackage{amssymb}
\usepackage{amsmath}
\usepackage{graphicx}
\usepackage{epsfig}
\usepackage{graphicx}
\usepackage{xcolor}
\usepackage{float}

\usepackage[
pagebackref=false,
colorlinks=true,
linkcolor=blue,
urlcolor=blue,
filecolor=black,
citecolor=blue,
pdfstartview=FitV,
pdftitle={},
pdfauthor={},
pdfsubject={},
pdfkeywords={},
pdfpagemode=None,
bookmarksopen=true
]{hyperref}

\begin{document}

\title{Integrability Breaking and Coherent Dynamics in Hermitian
	and Non-Hermitian Spin Chains with Long-Range Coupling}

\author{Y. S. Liu}
\affiliation{College of Physics and Materials Science, Tianjin Normal
	University, Tianjin 300387, China }

\author{X. Z. Zhang} \altaffiliation{zhangxz@tjnu.edu.cn}
\affiliation{College of Physics and Materials Science, Tianjin Normal
	University, Tianjin 300387, China }

\begin{abstract}
Unraveling the mechanisms of ergodicity breaking in complex quantum systems is a central pursuit in nonequilibrium physics. In this work, we investigate a one-dimensional spin model featuring a tunable long-range hopping term,	$H_{n}$, which introduces nonlocal interactions and bridges the gap between Hermitian and non-Hermitian regimes. Through a systematic analysis of level-spacing statistics, Krylov complexity, and entanglement entropy, we demonstrate that $H_{n}$ acts as a universal control parameter driving the transition from integrability to quantum chaos. Specifically, increasing the strength of $H_{n}$ induces a crossover from Poissonian to Gaussian Orthogonal Ensemble statistics in the Hermitian limit, and similarly triggers chaotic dynamics in the non-Hermitian case. Most remarkably, despite the onset of global chaos, we identify a tower of exact nonthermal eigenstates that evade thermalization. These states survive as robust quantum many-body scars, retaining low entanglement and coherent dynamics even under strong non-Hermitian perturbations. Our findings reveal a universal mechanism by which long-range and non-Hermitian effects reshape quantum ergodicity, offering new pathways for preserving quantum coherence in complex many-body systems.
\end{abstract}

\maketitle

\section{Introduction}
A central pursuit in modern quantum physics is to understand how quantum coherence can persist in complex, interacting many-body systems—an essential requirement for scalable quantum technologies and fundamental to nonequilibrium statistical mechanics \cite{np1,np2,np3,np4,np5,np6}. In generic isolated systems, coherence is rapidly lost as the system thermalizes, a process captured by the Eigenstate Thermalization Hypothesis (ETH) \cite{coll,eth,eth2,eth3,eth4ini,uto,web,evo,ZG1}. According to the ETH, each eigenstate behaves thermally with respect to local observables, implying that any initial state evolves toward a Gibbs ensemble and forgets its microscopic details. Yet, numerous mechanisms are now known to evade this fate. Integrable systems maintain an extensive set of conserved quantities that constrain their dynamics, while many-body localization (MBL) suppresses transport and thermalization through disorder-induced emergent integrals of motion \cite{inmbl1,inmbl2,inmbl3,inmbl4,inmbl5,inmbl6}. More recently, the discovery of quantum many-body scars (QMBS) has revealed a subtler route to ergodicity breaking \cite{qmbs1,qmbs2,qmbs3,qmbs4,qmbs5,qmbs6}. In these systems, a small subspace of atypical, low-entanglement eigenstates coexists within an otherwise thermal spectrum. When the system is initialized with finite overlap on this subspace, it exhibits long-lived oscillations and nonthermal dynamics despite the surrounding chaotic background.

However, truly isolated systems are an abstraction. In realistic settings, interaction with the environment—through dissipation, loss, or measurement—inevitably renders the dynamics non-Hermitian (NH) \cite{isnh1,isnh2,isnh3,isnh4}. Non-Hermitian Hamiltonians thus provide a powerful framework to model open quantum systems, capturing both gain–loss processes and conditional evolution \cite{isnh2,isnh5,isnh6}. This extension challenges conventional notions of thermalization and chaos: when the spectrum becomes complex, the familiar diagnostics of integrability based on real-level statistics cease to apply.

To address this, new spectral measures have been introduced, including the complex spacing ratio (CSR) \cite{CSR1,CSRSVD,CSRnhscar} and singular-value statistics \cite{CSRSVD,SVDstatistics}, which generalize random matrix theory (RMT) concepts to the complex plane. Yet, despite these advances, the interplay between non-Hermiticity, integrability breaking, and ergodicity remains only partially understood. In particular, the fate of quantum many-body scars in systems that are simultaneously non-Hermitian and nonintegrable remains an open and largely unexplored problem.

In this work, we address this challenge by studying a one-dimensional spin-1 model that includes a tunable long-range hopping term $H_{n}$. This term, controlled by a complex coupling $J_{n}$, introduces nonlocal interactions and provides a unified knob for driving the system from the Hermitian to the non-Hermitian regime. We demonstrate that $H_{n}$ acts as a universal integrability-breaking mechanism: in the Hermitian case, increasing $J_{n}$ induces a clear transition from Poissonian to Gaussian Orthogonal Ensemble (GOE) statistics, whereas in the non-Hermitian case, a complex $J_{n}$ produces an analogous crossover characterized by CSR and Krylov complexity. Our key finding is that a tower of exact, nonthermal eigenstates, originating from the integrable limit, survives both Hermitian and non-Hermitian perturbations. These states retain low entanglement and give rise to coherent fidelity revivals even deep in the chaotic regime. This persistence reveals a striking robustness of scarred subspaces against both nonlocal interactions and non-Hermitian deformations, suggesting a universal mechanism by which quantum coherence can persist beyond conventional integrable settings.

The remainder of this paper is organized as follows. Section \ref{sec:model} introduces the model Hamiltonian and its symmetry structure, and derives the exact ferromagnetic tower of eigenstates. Section \ref{sec:hermitian} analyzes the Hermitian regime, mapping the integrability-to-chaos transition using level statistics and Krylov complexity, and identifies the persistence of scars through entanglement and fidelity diagnostics. Section \ref{sec:non-hermitian} extends this analysis to the non-Hermitian regime, employing CSR and dynamical measures to confirm a parallel transition and the survival of the scar states. Section \ref{sec:summary} concludes with a summary and outlook. Technical derivations and computational methods are provided in Appendices \ref{Proof that the Tower States are Eigenstates} and \ref{Bi-Lanczos Algorithm for Non-Hermitian Systems}.

\section{Model Hamiltonian}
\label{sec:model}
Tremendous efforts have been devoted to exploring the transition from
integrable to non-integrable dynamics in quantum many-body systems,
particularly in spin-$1/2$ chains, where the breaking of integrability is
typically induced by disorder or kinetic constraints \cite{ZG1,q2,spin121,spin122,spin123,spin124,spin125,spin126}. In contrast, here we
demonstrate that a spin-$1$ chain with long-range spin-orbital coupling can
host a tower of many-body scar states residing in arbitrary energy sectors
modulated by an external magnetic field. The model Hamiltonian is given by 
\begin{equation}
H=J_{h}H_{h}+J_{c}H_{c}+J_{z}H_{z}+\sum_{n}J_{n}H_{n},  \label{H_full}
\end{equation}
where $H_{h}$, $H_{c}$, $H_{z}$, and $H_{n}$ denote Hermitian operators
describing distinct physical interactions. The coefficients $J_{h}$, $J_{c}$
, $J_{z}$, and $J_{n}$ characterize the relative strength and phase of the
corresponding processes. Their complex nature implies that the Hermiticity
of the full Hamiltonian is not guaranteed and depends sensitively on
microscopic control parameters.

The first term 
\begin{equation}
H_{h}=\sum_{j=1}^{N}\mathbf{S}_{j}\cdot \mathbf{S}_{j+1}
\end{equation}
represents the spin-$1$ Heisenberg exchange interaction. At each lattice
site $j$ the spin operators $\mathbf{S}_{j}=(S_{j}^{x},S_{j}^{y},S_{j}^{z})$
satisfy the usual SU$(2)$ commutation relations 
\begin{equation}
\left[ S_{j}^{\alpha },S_{k}^{\beta }\right] =i\delta _{jk}\sum_{\gamma
}\varepsilon _{\alpha \beta \gamma }S_{j}^{\gamma },\qquad \alpha ,\beta
,\gamma \in \{x,y,z\},
\end{equation}
where $\varepsilon _{\alpha \beta \gamma }$ is the Levi-Civita symbol, defined as
\begin{equation}
	\varepsilon_{\alpha\beta\gamma}=\left\{
	\begin{array}
		[c]{ll}%
		+1, & \text{if }\left(  \alpha\beta\gamma\right)  =\left(  xyz\right)
		,\left(  yzx\right)  ,\left(  zxy\right)  \text{,}\\
		-1, & \text{if }\left(  \alpha\beta\gamma\right)  =\left(  xzy\right)
		,\left(  yxz\right)  ,\left(  zyx\right)  \text{,}\\
		0, & \text{if any two indices are equal,}%
	\end{array}
	\right.
\end{equation}. The ladder operators are defined in
the standard manner: 
\begin{eqnarray}
S_{j}^{\pm } &\equiv &S_{j}^{x}\pm iS_{j}^{y}, \\
\left[ S_{j}^{z},S_{k}^{\pm }\right] &=&\pm \delta _{jk}S_{j}^{\pm }, \\
\left[ S_{j}^{+},S_{k}^{-}\right] &=&2\delta _{jk}S_{j}^{z}.
\end{eqnarray}
In the $S^{z}$-basis the single-site eigenstates are labeled by $m_{j}\in
\{+1,0,-1\}$ so that $S_{j}^{z}|m_{j}\rangle =m_{j}|m_{j}\rangle $ and $%
S_{j}^{+}|+1\rangle =0$, $S_{j}^{-}|-1\rangle =0$. The global (total) spin
operators are 
\begin{equation}
S^{\pm }\equiv \sum_{j=1}^{N}S_{j}^{\pm },\qquad
S^{z}=\sum_{j=1}^{N}S_{j}^{z}.
\end{equation}
Experimentally, this term can be realized in ultracold atomic gases confined
in optical lattices, where effective spin-1 degrees of freedom arise from
hyperfine manifolds of alkali atoms such as $^{23}$Na or $^{87}$Rb \cite{Heiexperiment1,Heiexperiment2,Heiexperiment3}. The superexchange interaction between neighboring
lattice sites gives rise to the effective coupling $J_{h}$, whose magnitude
and sign can be tuned via optical lattice depth and spin-dependent
collisions \cite{Heisuperexchange1,Heisuperexchange2,Heisuperexchange3}.

The second term describes a chiral three-spin interaction, 
\begin{equation}
H_{c}=\sum_{j=1}^{N}\mathbf{S}_{j}\cdot (\mathbf{S}_{j+1}\times \mathbf{S}%
_{j+2}),
\end{equation}%
which breaks both time-reversal and inversion symmetries
\cite{subspacescar}. Such a chiral coupling can emerge in spin-orbit-coupled
Mott insulators through higher-order superexchange processes or be
engineered synthetically in Rydberg-atom arrays and trapped-ion platforms
via Floquet modulation schemes that imprint complex tunneling phases \cite{chiralexperiment1,chiralexperiment2,chiralexperiment3,chiralexperiment4}. The strength $J_{c}$ can thus be dynamically tuned
by adjusting the driving amplitude or the relative phase between neighboring
bonds.

The third term 
\begin{equation}
H_{z}=\sum_{j=1}^{N}S_{j}^{z}
\end{equation}
is the Zeeman coupling between the local spin and an external magnetic field
along the $z$ direction. This term can be directly controlled by the applied
magnetic field strength or by effective AC-Stark shifts in optical or
microwave dressing schemes, enabling selective tuning of the on-site energy
splitting \cite{ACStark}.

Finally, the nonlocal hopping term 
\begin{equation}
H_{n}=i\sum_{j=1}^{N}\left( S_{j}^{+}S_{j+n}^{-}-S_{j}^{-}S_{j+n}^{+}\right)
,  \label{H_n}
\end{equation}
represents a long-range spin-exchange process with hopping distance $n$,
where 
\begin{equation}
n= 
\begin{cases}
1,2,\cdots ,N/2-1, & N\text{ even,} \\ 
1,2,\cdots ,(N-1)/2, & N\text{ odd,}%
\end{cases}%
\end{equation}
under periodic boundary conditions (PBC). The term $H_{n}$ naturally arises in systems with long-range dipole--dipole or spin--orbit-mediated interactions, such as polar molecules, trapped ions, or cavity-QED arrays. In such platforms, the coupling strength $J_{n}$ can be rendered complex by controlling laser-induced Raman transitions, thereby introducing tunable synthetic gauge phases. In particular, trapped-ion quantum simulators have demonstrated laser-mediated spin couplings whose interaction range scales inversely with system size, successfully engineering long-range effective interactions \cite{longrange}.

\subsection{Symmetries, and Exact Ferromagnetic States}

\subsubsection{Symmetries}

We now summarize the symmetry structure of $H$. (i) U(1) symmetry: Each term
in $H$ commutes with the total magnetization $S^{z}=\sum_{j}S_{j}^{z}$,
i.e., $[S^{z},H_{k}]=0$ for $k=h,$ $c,$ $z,$ $n$. Thus $S^{z}$ (or
equivalently the total magnetization $M\equiv S^{z}$) is a constant of
motion and the full Hilbert space decomposes into orthogonal magnetization
sectors 
\begin{equation}
H=\bigoplus_{M=-N}^{+N}\mathcal{H}_{M},
\end{equation}%
where%
\begin{equation}
\mathcal{H}_{M}\equiv \{\,\left\vert \psi \left( p\right) \right\rangle
:S^{z}\left\vert \psi \left( p\right) \right\rangle =M\left\vert \psi \left(
p\right) \right\rangle \,\}.
\end{equation}%
We will often parametrize a sector by the number of spin-lowering operations 
$p$ relative to the fully polarized state (see below), so that $M=N-p$, where $p=0,1,2,...,2N$. (ii)
Lattice translation symmetry: Assuming uniform couplings and PBC,
the Hamiltonian is invariant under the one-site translation operator $T$.
Hence, crystal momentum $k$ is a good quantum number, and each sector $%
\mathcal{H}_{M}$ further splits into $\mathcal{H}_{M}=\bigoplus_{k}\mathcal{H%
}_{M,k}$, where $T\left\vert \psi _{M,k}\right\rangle =e^{ik}\left\vert \psi
_{M,k}\right\rangle $. (iii) Global SU(2) symmetry: The Heisenberg term $%
H_{h}$ and the chiral term $H_{c}$ are rotational scalars, satisfying $[%
\mathbf{S}^{2},H_{h/c}]=0$ and $[S^{\alpha },H_{h/c}]=0$ $(\alpha =x,y)$ \cite{subspacescar,ZG1}. However,
the subgroup generated by $S^{z}$ an term $H_{z}$ and the nonlocal term $%
H_{n}$ break the full SU(2) symmetry down to the $U(1)$ . Consequently, the
total spin $\mathbf{S}^{2}$ is not a conserved quantity for the generic
Hamiltonian.

\subsubsection{Symmetry-determined subspaces}

Collecting the results above, the Hamiltonian block-diagonalizes into
symmetry-labeled subspaces. A convenient labeling is 
\begin{equation}
H=\bigoplus_{M=-N}^{N}\bigoplus_{k}\mathcal{H}_{M,k},
\end{equation}%
where $M$ (total $S^{z}$) and $k$ (crystal momentum) are exact quantum
numbers for the full Hamiltonian defined above. If additionally $%
J_{z}=J_{n}=0$ (pure SU(2)-symmetric point), each $(M,k)$ block further
decomposes into irreducible $\mathrm{SU}(2)$ multiplets labeled by total
spin $S_{\mathrm{tot}}$ (eigenvalues of $\mathbf{S}^{2}$). If $J_{c}=0$ and
all coefficients are real, spatial parity may also be a symmetry and can be
used to refine the block decomposition; the presence of $H_{c}$ usually
removes parity as a good quantum number.

\subsubsection{Exact ferromagnetic tower of states}

Despite the complex structure of $H$, it possesses an exactly solvable tower
of ferromagnetic states. These states are the totally symmetric
(zero-momentum) multiplet generated from the fully polarized highest-weight
state $\left\vert \Uparrow\right\rangle \equiv\bigotimes_{j=1}^{N}%
|+1\rangle_{j}$:
\begin{equation}
	\left\vert \psi\left(  N,N-p\right)  \right\rangle \equiv\frac{1}{\Omega
	}(S^{-})^{p}\left\vert \Uparrow\right\rangle \label{eq:tower}%
\end{equation}
where\ $\Omega=\sqrt{\left(  2N\right)  !p!/\left(  2N-p\right)  !}$. We can find their eigenenergies by evaluating the
action of each term on this tower. First, the highest-weight state $\left\vert
\Uparrow\right\rangle $ is a trivial eigenstate of all four terms:
$H_{h}\left\vert \Uparrow\right\rangle =N\left\vert \Uparrow\right\rangle $
(since each neighboring pair of fully aligned spin-1 moments satisfies
$S_{j}\cdot S_{j+1}|\Uparrow\rangle=1$, corresponding to a normalized
contribution of $+1$ per bond), $H_{c}\left\vert \Uparrow\right\rangle =0$ (as
the scalar triple product of parallel vectors vanishes), $H_{z}\left\vert
\Uparrow\right\rangle =N\left\vert \Uparrow\right\rangle $, and $H_{n}%
\left\vert \Uparrow\right\rangle =0$ (due to $S_{j}^{+}|+1\rangle_{j}=0$).

Because $H_{h}$ and $H_{c}$ are rotational scalars, they commute with $S^{-}$
and thus act invariantly on the entire tower, yielding $H_{h}\left\vert \psi
\left( N,N-p\right) \right\rangle =N\left\vert \psi \left( N,N-p\right)
\right\rangle $ and $H_{c}\left\vert \psi \left( N,N-p\right) \right\rangle
=0$ for all $p$. The Zeeman term acts as $H_{z}\left\vert \psi \left(
N,N-p\right) \right\rangle =(N-p)\left\vert \psi \left( N,N-p\right)
\right\rangle $, which follows from $[S^{z},(S^{-})^{p}]=-p\,(S^{-})^{p}$.
Finally, the nonlocal term $H_{n}$ annihilates every state in the tower, $%
H_{n}\left\vert \psi \left( N,N-p\right) \right\rangle =0$. This arises
because $\left\vert \psi \left( N,N-p\right) \right\rangle $ is the $k=0$
(totally symmetric) $p$-magnon state, and the antisymmetric structure of the 
$H_{n}$ operator [Eq.~(\ref{H_n})] results in a vanishing eigenvalue for
this uniform superposition (see Appendix \ref{Proof that the Tower States
are Eigenstates} for more details).

Combining these results, we find that the states $\left\vert \psi \left(
N,N-p\right) \right\rangle $ are exact eigenstates of the full Hamiltonian
[Eq.~(\ref{H_full})] with energies 
\begin{eqnarray}
H\left\vert \psi \left( N,N-p\right) \right\rangle &=&E(p)\left\vert \psi
\left( N,N-p\right) \right\rangle , \\
E(p) &=&J_{h}N+J_{z}(N-p).  \label{eq:tower-energy}
\end{eqnarray}%
This tower of exact states spans all magnetization sectors $M = N - p$, featuring an energy spectrum that varies linearly with $M$. Remarkably, these states remain robust even in the presence of the non-integrable interactions $H_{c}$ and $H_{n}$.

Here two points should be highlighted: 1. The crucial property that makes
the states Eq. (\ref{eq:tower}) exactly solvable is that they are the totally
symmetric (zero-momentum) multiplet generated from the highest-weight state $%
\left\vert \Uparrow \right\rangle $. Any Hamiltonian that (i) acts as a
scalar on this multiplet (e.g. SU(2)-invariant two- and three-spin scalar
operators) and (ii) annihilates the uniform magnon superposition (as $H_{n}$
does in the antisymmetric form above) will preserve the tower. 2. If one
perturbs the model by terms that break translation invariance (disorder) or
introduce momentum-dependent hopping that does not annihilate the $k=0$
magnon, the tower generically hybridizes with other states and ceases to be
exact; nevertheless, for sufficiently weak perturbations the tower may
survive as long-lived scarred eigenstates (quasi-exact).

In the following sections, we will explore the spectral and dynamical properties of this model and demonstrate how a tower of nonthermal eigenstates persists despite the presence of non-integrable, long-range interactions. For simplicity, in the subsequent analysis, we focus on the case with a single $n = 3$ term. Nevertheless, calculations performed with the full Hamiltonian in Eq.~(\ref{H_full}), which includes the summation over all $H_{n}$, yield qualitatively similar results.

\section{Spectral Statistics and Quantum Scar in the Hermitian Limit}
\label{sec:hermitian}
\subsection{Crossover between the intrgrablity and chaos}

Before investigating the emergence of quantum many-body scars and the
associated revival dynamics, it is essential to first characterize the
integrability of the system in its Hermitian limit. We focus on the case
where all coupling constants $J_{h}$, $J_{c}$, and $J_{n}$ are real. In this
regime, the Hamiltonian remains Hermitian, allowing us to quantify
integrability through spectral statistics.

\begin{figure}[tbp]
	\centering
	\includegraphics[width=0.9\linewidth]{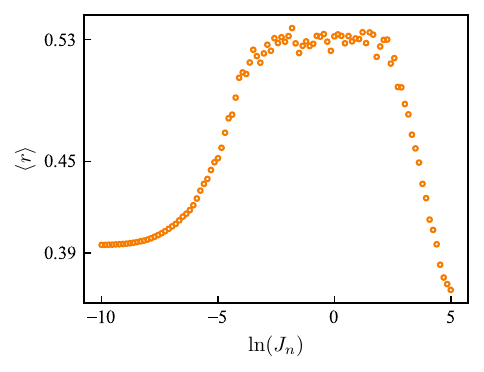}
	\caption{Dependence of the average level spacing ratio $\langle r\rangle$ on the nonlocal hopping strength $J_{n}$. 
		Calculations are performed for a system with parameters $(J_{h}, J_{c}, J_{z}, n) = (1, 1, 0.5, 3)$ under PBC. 
		As $J_{n}$ increases, the system exhibits a clear transition from an integrable regime to a chaotic one, and eventually reenters the integrable regime at large $J_{n}$. 
		Results correspond to a system size of $N = 12$, within the $(M, k) = (0, 0)$ subspace whose Hilbert-space dimension is $6166$.}
	\label{fig:herm_Jn_and_r}
\end{figure}
To this end, we analyze the nearest-neighbor level spacings of the Hamiltonian
eigenenergies. Denoting the ordered spectrum as $\{E_{i}\}$, we define the
level spacing as $s_{i}=E_{i+1}-E_{i}$. The statistical distribution of
$\{s_{i}\}$ serves as a powerful diagnostic for quantum integrability. A
particularly robust and widely used indicator is the ratio of adjacent level
spacings,
\begin{equation}
	r_{i}=\min\!\left(  \frac{s_{i-1}}{s_{i}},\frac{s_{i}}{s_{i-1}}\right)
	,\label{mean_level_spacing}%
\end{equation}
whose ensemble average $\langle r\rangle$ characterizes the underlying
spectral correlations \cite{meanlevelspacing}. In integrable systems, energy
levels are uncorrelated and follow a Poisson distribution, yielding $\langle
r\rangle_{\mathrm{Poisson}}\approx0.386$. By contrast, non-integrable
(chaotic) systems display level repulsion consistent with Wigner-Dyson
statistics, where $\langle r\rangle_{\mathrm{GOE}}\approx0.536$ or $\langle
r\rangle_{\mathrm{GUE}}\approx0.603$, depending on the symmetry class~of the
system \cite{subspacescar,meanlevelspacing}.

\begin{figure}[tbh]
	\centering\includegraphics[width=0.9\linewidth]{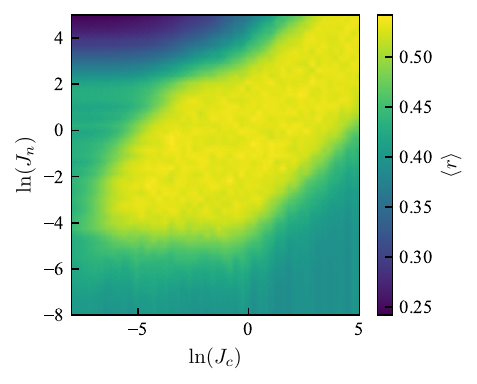}
	\caption{
		Phase diagram in the plane of chiral coupling $J_{c}$ and nonlocal coupling
		$J_{n}$. The color map shows the mean level-spacing ratio $\langle r\rangle$
		computed within the $\mathcal{H}_{M=0,k=0}$ symmetry sector for
		$(J_{h},J_{z},n)=(1,0.5,3)$. The system remains integrable when either
		$J_{c}$ or $J_{n}$ is weak, whereas clear signatures of quantum chaos appear
		only when both couplings become comparable in magnitude. Data shown are for a system size $N=12$, corresponding to a Hilbert-space
		dimension of $6166$.
	}
	\label{fig:herm_Jc_Jn_r}
\end{figure}
To identify the role of the nonlocal term $H_{n}$, we compare two cases: 
\begin{align}
H_{0}& =J_{h}H_{h}+J_{c}H_{c}+J_{z}H_{z}, \\
H& =H_{0}+J_{n}H_{n}.
\end{align}
The comparison between $H_{0}$ and $H$ allows us to isolate the effect of $%
H_{n}$ on integrability. We compute the mean spacing ratio $\langle r\rangle 
$ as a function of $J_{n}$, within the symmetry-resolved subspace $\mathcal{%
H }_{M=0,k=0}$. As shown in Fig.~\ref{fig:herm_Jn_and_r},
the system undergoes a crossover from integrability to quantum chaos and
then back to near-integrability as $J_{n}$ increases from zero. This
nonmonotonic behavior highlights the delicate balance between local and
nonlocal spin interactions in shaping the dynamical complexity of the system.

To provide a comprehensive understanding of the model, we systematically analyze 
two key aspects: the phase diagram of the chiral coupling $J_{c}$ and the nonlocal coupling $J_{n}$ 
in terms of the average level spacing ratio $\langle r \rangle$. \begin{figure}[tbh]
	\centering\includegraphics[width=0.9\linewidth]{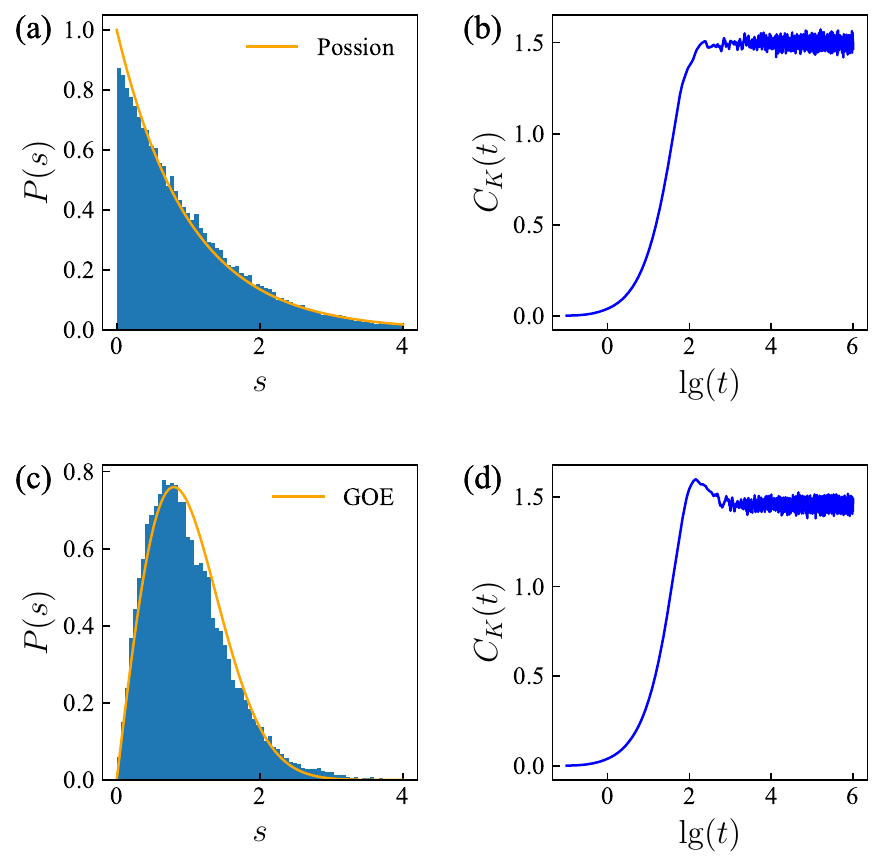}
	\caption{
		Distribution of nearest-neighbor level spacings and the corresponding average
		ratio $\langle r\rangle$ for the central part of the spectrum, together with
		the time evolution of the Krylov complexity of $\mathcal{H}_{M=0,k=0}$ under
		periodic boundary conditions. Calculations are performed with parameters
		$(J_{h},J_{c},J_{z},n)=(1,1,0.5,3)$. For $J_{n}=0$, panel~(a) shows a Poissonian level-spacing distribution with
		$\langle r\rangle\!\approx\!0.392$, indicative of integrability, and
		panel~(b) displays the Krylov complexity, which rapidly rises and
		saturates—characteristic of integrable dynamics. For $J_{n}=0.2$, panel~(c) exhibits a Gaussian-orthogonal-ensemble (GOE)
		distribution with $\langle r\rangle\!\approx\!0.529$, signaling the onset of
		quantum chaos, while panel~(d) shows a non-monotonic evolution of the
		Krylov complexity—initial growth, decay, and eventual saturation—typical of
		chaotic systems. These results together demonstrate a clear transition from integrability to
		chaos as $J_{n}$ increases, consistent with the mean level-spacing ratio
		defined in Eq.~(\ref{mean_level_spacing}). Panels~(a) and (c) correspond to a
		system size of $N=14$ (Hilbert-space dimension $44046$), while panels~(b) and
		(d) correspond to $N=10$ (dimension $902$).
	}
	\label{fig:herm_krylov_energy}
\end{figure}
The results, presented in Fig.~\ref{fig:herm_Jc_Jn_r}, 
demonstrate that within the studied subspace $(M,k)=(0,0)$, 
the system remains in the integrable regime when both coupling parameters $J_{c}$ and $J_{n}$ are weak. 
As $J_{c}$ and $J_{n}$ increase, the system gradually transitions from an integrable to a non-integrable regime. 
Notably, when both couplings become sufficiently strong, the contribution of $H_{h}$ becomes negligible, 
yet the system still exhibits non-integrable (chaotic) behavior. 
This finding highlights that the interplay between $H_{c}$ and $H_{n}$ 
plays a dominant role in determining the transition between integrability and chaos in this model. For $J_{n}=0$, the Hamiltonian $H_{0}$ corresponds to an integrable model, 
as evidenced by the Poissonian level-spacing statistics. 
Introducing a finite, yet small, $J_{n}$ breaks integrability and drives the system toward the Wigner–Dyson regime, 
as illustrated by the numerical results shown in panels (a) and (c) of Fig.~\ref{fig:herm_krylov_energy}.

This transition serves as the foundation for understanding how weak nonlocal
perturbations generate scar-like coherent dynamics in an otherwise chaotic
spectrum.

Beyond static spectral diagnostics, the integrable-to-chaotic transition can
be further characterized from a dynamical perspective. We employ the Krylov
complexity framework, which quantifies how an initial quantum state spreads
under repeated action of the Hamiltonian. Operationally, the Krylov basis is
constructed through iterative applications of $H$ followed by
orthogonalization with respect to the previously generated basis vectors \cite{krylovcomplexity1,krylovcomplexity2,krylovcomplexity3,krylovcomplexity4,krylovcomplexity5}. To
ensure consistency between the Hermitian and non-Hermitian analyses, we
adopt the bi-Lanczos algorithm for numerical implementation. Details of this algorithm are
provided in Appendix~\ref{Bi-Lanczos Algorithm for Non-Hermitian Systems}.
This approach enables us to capture both spectral and dynamical signatures
of integrability breaking and to connect them directly to the emergence of
many-body scars and revival dynamics.

The bi-Lanczos algorithm maps the time evolution of a quantum state under a
(possibly non-Hermitian) Hamiltonian onto an effective single-particle
tight-binding model defined on a semi-infinite chain \cite{krylovcomplexity1}. This mapping provides an intuitive and
powerful framework for analyzing the complex dynamical behavior of quantum
systems and serves as an efficient diagnostic tool to distinguish between
integrable and chaotic regimes.

To construct the Krylov basis for a non-Hermitian Hamiltonian $H$, one
introduces two sets of biorthogonal basis vectors $\{\lvert q_{n}\rangle \}$
and $\{\lvert p_{n}\rangle \}$ satisfying 
\begin{align}
	\langle p_{m}|q_{n}\rangle & =\delta_{mn},\,\sum_{n}|q_{n}\rangle\langle
	p_{n}|=1,\\
	\lvert q_{0}\rangle & =\lvert p_{0}\rangle=\lvert\psi(0)\rangle.
\end{align}
The subsequent Krylov vectors are generated recursively through the
bi-Lanczos iteration: 
\begin{align}
H\lvert q_{n}\rangle & =a_{n}\lvert q_{n}\rangle +b_{n+1}\lvert
q_{n+1}\rangle +c_{n}\lvert q_{n-1}\rangle , \\
H^{\dagger }\lvert p_{n}\rangle & =a_{n}^{\ast }\lvert p_{n}\rangle
+c_{n+1}^{\ast }\lvert p_{n+1}\rangle +b_{n}^{\ast }\lvert p_{n-1}\rangle ,
\end{align}%
where the coefficients $a_{n}$, $b_{n}$, and $c_{n}$ are determined so that
the biorthogonality condition is maintained at each step.

In this Krylov representation, the time-dependent wave function can be
expanded as 
\begin{equation}
\lvert \psi (t)\rangle =\sum_{n}\phi _{n}(t)\lvert q_{n}\rangle ,
\end{equation}
where $\phi _{n}(t)=\langle p_{n}|\psi (t)\rangle $ denotes the projection
of the evolving state onto the $n$-th Krylov vector. The amplitudes $\phi
_{n}(t)$ satisfy a discrete Schr\"{o}dinger-like equation that resembles a
single-particle tight-binding model on a semi-infinite chain: 
\begin{equation}
i\frac{\partial }{\partial t}\phi _{n}(t)=b_{n}\phi _{n-1}(t)+a_{n}\phi
_{n}(t)+c_{n+1}\phi _{n+1}(t).  \label{krylov_eom}
\end{equation}
Here $\{a_{n},b_{n},c_{n}\}$ are the Lanczos coefficients obtained from the
bi-Lanczos recursion described in Appendix~\ref{Algorithmic Construction}.
This recurrence relation is mathematically equivalent to a tight-binding
chain where $a_{n}$ represents on-site energies and $b_{n}$, $c_{n}$ denote
the hopping amplitudes between neighboring sites.

Once the coefficients $\phi _{n}(t)$ are obtained, the Krylov complexity is
defined as \cite{krylovcomplexity1,krylovcomplexity2,krylovcomplexity3,krylovcomplexity4,krylovcomplexity5}
\begin{equation}
C_{K}(t)=\sum_{n}n\,|\phi _{n}(t)|^{2},  \label{krylov_complexity}
\end{equation}
which quantifies the spread of the quantum state in the Krylov space, i.e.,
how far the evolved state has propagated along the effective chain.

For Hermitian and non-Hermitian systems, the behavior of $C_{K}(t)$ provides a clear dynamical
signature of integrability. In integrable systems, $C_{K}(t)$ rapidly
saturates to a steady plateau, reflecting quasi-periodic and restricted
motion in Krylov space. In contrast, for chaotic systems, $C_{K}(t)$
exhibits a rapid initial growth followed by saturation at a much
higher value, signaling extensive spreading across the Krylov basis \cite{krylovcomplexity2}.

Next, within the subspace $(M,k)=(0,0)$, we compute the Krylov complexity for
various values of $J_{n}$, using as the initial state the equal-weight linear
superposition of the eigenstates of the Hermitian matrix obtained by taking
the lower-triangular part of the Hamiltonian and restricting it to this
subspace. As shown in panels (b) and (d) of Fig.~\ref{fig:herm_krylov_energy}, for $J_{n}=0$, the Krylov complexity
quickly reaches a steady plateau, consistent with integrable dynamics.
Conversely, for $J_{n}=0.2$, $C_{K}(t)$ displays a pronounced growth before
saturation, indicative of chaotic evolution. These contrasting behaviors
demonstrate that Krylov complexity serves as a sensitive probe distinguishing
integrable and chaotic regimes within the same microscopic model.

\subsection{The evidence of the Quantum many-body scars}

While the introduction of the nonlocal hopping term $H_{n}$ drives the
system from integrability to chaos, a set of nonthermal states originating
from the tower states $\{|\psi_{n}(N,N-p)\rangle\}$ can persist deep in the
chaotic regime. To understand this phenomenon, it is instructive to first
discuss the underlying structure of the Hilbert space.

\begin{figure}[tbp]
	\centering
	\includegraphics[width=0.9\linewidth]{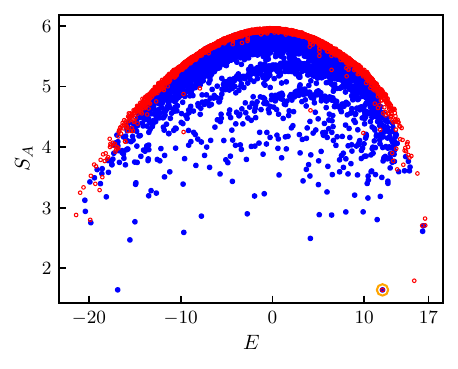}
	\caption{Von Neumann entanglement entropy, computed using Eq.~(\ref{ent_entropy}), for
		a system of size $N=12$ under periodic boundary conditions (PBC) with
		parameters $(J_{h},J_{c},J_{z},n)=(1,1,0.5,3)$. Within the $(M,k)=(0,0)$
		subspace (Hilbert-space dimension is 6166), the entanglement entropies of all
		eigenstates are plotted for $J_{n}=0$ (blue dots) and $J_{n}=0.2$ (red
		circles). A notable feature is the persistence of a low-entanglement state in
		the lower-right region (an orange circle) across the integrable-to-chaotic
		transition, indicating its robustness as a quantum scar state.}
	
	\label{fig:herm_entropy}
\end{figure}
In our spin-$1$ model, the conservation of the total magnetization $S_{%
\mathrm{tot}}^{z}=\sum_{j}S_{j}^{z}$ divides the full Hilbert space into
dynamically disconnected subspaces labeled by the quantum number $M=S_{%
\mathrm{tot}}^{z}$. This fragmentation implies that the Hamiltonian is
block-diagonal, $H=\bigoplus_{M}\mathcal{H}_{M}$, and each block $\mathcal{H}%
_{M}$ evolves independently under unitary dynamics. Within each fixed-$M$
sector, the tower state Eq. (\ref{eq:tower}) resides as a highly
symmetric state with minimal entanglement. These states remain exact eigenstates of the Hamiltonian even when $J_{n}\neq 0$, 
demonstrating their robustness against the non-integrable perturbations introduced by $H_{n}$. 
Despite the presence of chaotic dynamics in the remaining spectrum, these exact nonthermal states coexist 
with thermal eigenstates, a characteristic feature reminiscent of quantum many-body scars.

\begin{figure}[tbp]
	\centering
	\includegraphics[width=0.9\linewidth]{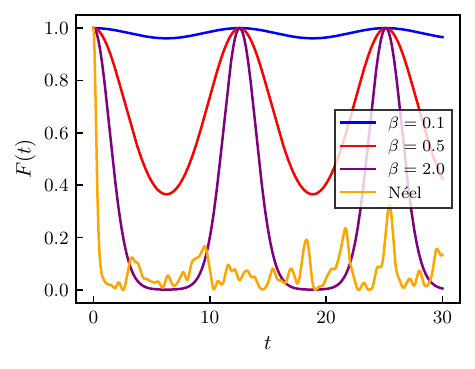}
	\caption{Fidelity dynamics of the Hamiltonian $H$ with parameters
		$(J_{h},J_{c},J_{z},J_{n},n)=(1,1,0.5,0.2,3)$ under periodic boundary
		conditions, for a system size of $N=8$ with a Hilbert-space dimension of
		$6561$. When the initial state is chosen as the coherent state defined in
		Eq.~(\ref{psibeta}), clear periodic revivals are observed, indicating that
		this coherent state, or equivalently, the tower of states defined in
		Eq.~(\ref{eq:tower}), exhibits the characteristic behavior of a quantum
		many-body scar. In contrast, when initialized in the N\'{e}el state, the
		fidelity rapidly decays to zero, signaling fast thermalization and the
		absence of scar dynamics.
	}
	
	\label{fig:herm_fidelity}
\end{figure}
To characterize the nonthermal nature of these states, we compute the
bipartite entanglement entropy of each eigenstate $|\psi _{n}\rangle $. The
system is divided into two subsystems $A$ and $B$, and for each eigenstate,
we construct the reduced density matrix of subsystem $A$ as 
\begin{equation}
\rho _{A,n}=\mathrm{Tr}_{B}\!\left( |\psi _{n}\rangle \langle \psi
_{n}|\right) ,
\end{equation}
and define the von Neumann entanglement entropy 
\begin{equation}
S_{A}=-\mathrm{Tr}\!\left[ \rho _{A,n}\ln \rho _{A,n}\right] .
\label{ent_entropy}
\end{equation}
As shown in Fig.~\ref{fig:herm_entropy}, the entropy distribution in the chaotic regime displays a sharp dip
corresponding to one particular eigenstate whose entanglement remains
anomalously low. This state can be directly associated with one member of
the tower $\{|\psi _{n}(N,N-p)\rangle \}$, demonstrating that the tower
survives as a scarred subspace even after the system becomes globally
nonintegrable.

\begin{figure*}[htbp]
	\centering
	\includegraphics[width=0.9\linewidth]{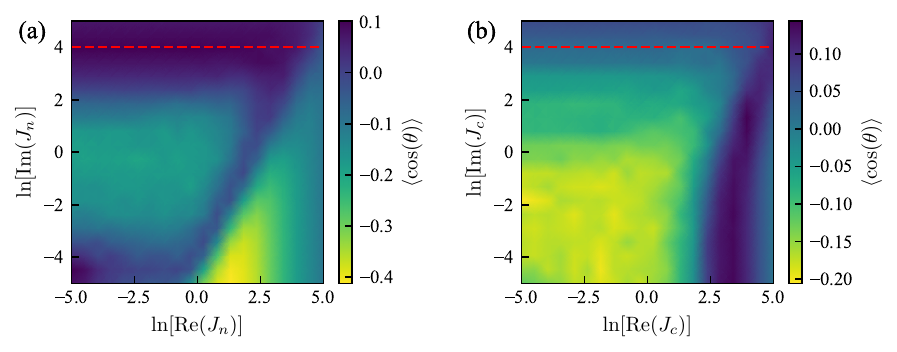}
	\caption{Phase diagrams of the non-Hermitian spin system under complex coupling 
		parameters (a)~$J_{n}$ and (b)~$J_{c}$. 
		(a)~With $J_{c}$ fixed at $1$, the system exhibits an 
		integrable--nonintegrable--integrable crossover as either the real or 
		imaginary component of $J_{n}$ is increased while the other component is held 
		constant. Notably, when the imaginary part of $J_{n}$ becomes sufficiently 
		large, e.g., $\ln[\operatorname{Im}(J_{n})]=4$, variations in its real part no 
		longer drive a transition between integrable and chaotic behavior. 
		(b)~With $J_{n}$ fixed at $0.2i$, the system remains in the chaotic regime for 
		small $|J_{c}|$. In contrast, a sufficiently large imaginary component of $J_{c}$ 
		suppresses the transition toward integrability, as indicated by the red line. 
		For both panels, the parameters $(J_{h},J_{z},n)=(1,0.5,3)$, system size $N=12$, 
		and Hilbert-space dimension $6166$ are kept fixed.
	}
	\label{fig:nh_diagram}
\end{figure*}
To further probe the dynamical manifestation of these scarred states, we
construct a coherent superposition of the tower states as 
\begin{equation}
|\beta \rangle =e^{\beta S^{-}}|\Uparrow \rangle =\sum_{p=0}^{2N}\frac{\beta
^{p}}{p!}|\psi _{n}(N,N-p)\rangle ,  \label{psibeta}
\end{equation}
and study its fidelity dynamics under time evolution, 
\begin{equation}
F(t)=|\langle \beta |e^{-iHt}|\beta \rangle |^{2}.  \label{fidelity}
\end{equation}
As shown in Fig.~\ref{fig:herm_fidelity}, $F(t)$ exhibits pronounced periodic revivals for several choices of $\beta $
, signaling coherent oscillations characteristic of quantum many-body scars.
In contrast, a typical product state such as the N\'{e}el state, 
\begin{equation}
|\mathrm{N\acute{e}el}\rangle =\bigotimes_{j=1}^{N/2}(|+1\rangle\otimes|-1\rangle) ,  \label{neel}
\end{equation}
shows rapid relaxation and the absence of revival, consistent with generic
thermalization. Therefore, the tower states $\{|\psi _{n}(N,N-p)\rangle \}$
and their linear combinations constitute a scarred subspace that preserves
coherent revivals even when the surrounding spectrum exhibits chaotic
statistics.

\section{Extension to Non-Hermitian Systems}
\label{sec:non-hermitian}

In this section, we focus on the non-Hermitian regime where the 
coupling strength is complex, and investigate quantum many-body scars within
this framework. Similar to the Hermitian case, we first need to verify
whether the system is non-integrable under non-Hermitian conditions. Here,
since the Hamiltonian is non-Hermitian, its eigenvalue spectrum becomes
complex. To characterize chaotic behavior in this regime, we employ the
complex spacing ratio (CSR) analysis \cite{CSR1,CSRSVD,CSRnhscar}. For a non-Hermitian
Hamiltonian with complex eigenvalues $\{E_{j}\}$, we define
\begin{equation}
\lambda _{j}=\frac{E_{j}^{NN}-E_{j}}{E_{j}^{NNN}-E_{j}},
\label{CSR}
\end{equation}
where $E_{j}^{NN}$ and $E_{j}^{NNN}$ denote the nearest and
next-to-nearest neighboring eigenvalues of $E_{j}$, respectively. 
By construction, $|\lambda _{j}|\leqslant 1$, and $\lambda _{j}$ 
distributes inside the unit disk in the complex plane. 

\begin{figure}
	\centering
	\includegraphics[width=0.9\linewidth]{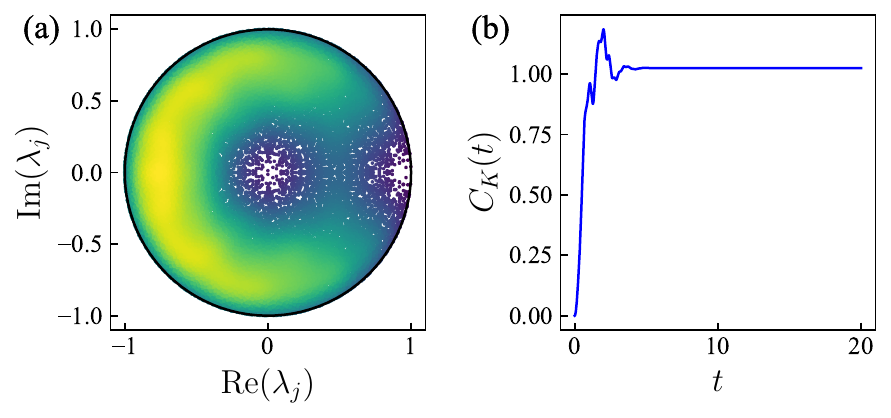}
	\caption{Spectral distribution and Krylov complexity of the non-Hermitian Hamiltonian 
		$\mathcal{H}_{M=0,k=0}$ with parameters $(J_{h},J_{c},J_{n},n)=(1,1,0.2i,3)$. 
		Panel~(a) shows the complex spacing ratio (CSR) statistics. The eigenvalues form 
		a sparse distribution in the complex plane, yielding 
		$\langle \cos\theta_{j} \rangle \approx -0.192$, characteristic of nonintegrable 
		behavior. Panel~(b) displays the corresponding Krylov complexity dynamics, which 
		exhibits an initial rapid growth followed by decay and eventual saturation, a 
		signature of chaotic evolution in non-Hermitian systems. Together, these results 
		demonstrate that the introduction of a complex-valued long-range coupling $H_{n}$ 
		drives the system from integrability to chaos, consistent with the comparison to 
		the $J_{n}=0$ case in Fig.~\ref{fig:herm_krylov_energy}. 
		Panel~(a) corresponds to system size $N=14$ with Hilbert-space dimension $44046$, 
		while panel~(b) uses $N=10$ with Hilbert-space dimension $902$.}
	\label{fig:nh1_krylov_energy}
\end{figure}
Following the analogy with the Berry--Tabor and
Bohigas--Giannoni--Schmit conjectures, one expects that integrable systems
exhibit CSR statistics identical to those of uncorrelated complex levels,
while chaotic systems follow the Ginibre ensemble. In the integrable limit,
the eigenvalues are independent, and the reference level does not influence
its two nearest neighbors. Consequently, all ratios $\lambda _{j}$ have equal
probability, resulting in a uniform (flat) distribution over the
unit circle. \begin{figure*}[t]
	\centering
	\includegraphics[width=0.9\linewidth]{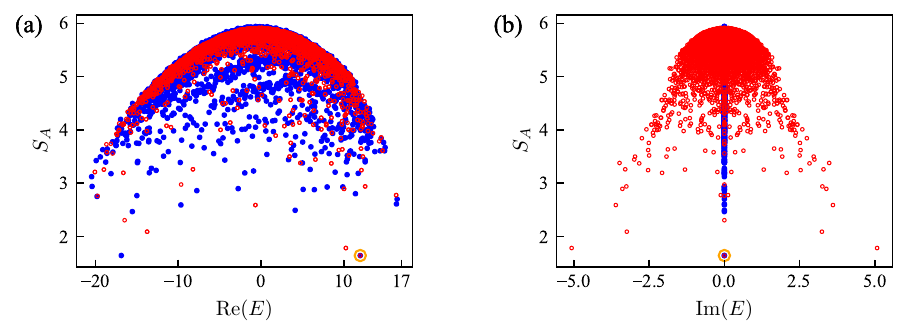}
	\caption{
		Von Neumann entanglement entropy computed using Eq.~(\ref{ent_entropy}) for the 
		right eigenvectors of $\mathcal{H}_{M=0,k=0}$ with parameters 
		$(J_{h},J_{c},J_{z})=(1,1,0.5)$ and $n=3$. The system size is $N=12$, and the 
		Hilbert-space dimension is $6166$. Panels~(a) and (b) show the dependence of the 
		entanglement entropy on the real and imaginary parts of the eigenvalues, 
		respectively. Blue dots correspond to the Hermitian case ($J_{n}=0$), while red 
		circles represent the non-Hermitian case ($J_{n}=0.2i$). A common low-entanglement 
		state appears in both parameter regimes (highlighted in orange), demonstrating 
		the robustness of the quantum many-body scar against non-Hermitian perturbations.
	}
	\label{fig:nh1_entropy}
\end{figure*}In contrast, for chaotic spectra obeying Ginibre statistics,
level repulsion emerges: the ratio density vanishes at the origin and is
suppressed at small angles, indicating the mutual repulsion of neighboring
levels. A quantitative indicator is provided by the angular average 
$\langle \cos \theta _{j}\rangle$, where $\theta_{j}=\arg\left(  \lambda_{j}\right)$, which approaches 
$\langle \cos \theta _{j}\rangle \approx 0$ for integrable systems and 
$\langle \cos \theta _{j}\rangle \approx -0.24$
for chaotic systems in the thermodynamic limit \cite{CSR1}. For finite-size systems, these values are approximated
but serve as clear diagnostics of the integrability-to-chaos transition.

In our previous study of the Hermitian case, we found that adjusting 
$J_{c}$ and $J_{n}$ can regulate the integrability of
the system. Therefore, in this non-Hermitian case, we primarily focus
on studying scenarios where $J_{c}$ and $J_{n}$ are
complex numbers. First, we calculate the CSR and the related 
$\langle \cos \theta_j \rangle$ using Eq.~(\ref{CSR}) to determine how the
integrable and chaotic characteristics vary with $J_{c}$ and $J_{n}$.
In Fig.~\ref{fig:nh_diagram}, we plot two phase diagrams that show the
relationship between $\langle \cos \theta_j \rangle$ and the other
parameter when fixing $J_{c}$ and $J_{n}$ individually. 

Building on this framework, we investigate two distinct scenarios: 
(i) the non-Hermitian transformation where introducing the complex term 
$H_{n}$ converts a Hermitian integrable system into a non-Hermitian chaotic system, 
and (ii) the regime where both $H_{c}$ and $H_{n}$ possess
complex-valued parameters, revealing that $H_{n}$ induces an
integrable-to-chaotic transition within non-Hermitian systems. 
We systematically analyze the quantum scarring characteristics in both cases,
with particular emphasis on their dynamical manifestations.

\subsection{\textit{Case I}: Complex $J_{n}$}

We now extend our analysis to non-Hermitian regimes by allowing the coupling 
$J_{n}$ to take complex values, while keeping $J_{h}$, $J_{c}$, and $J_{z}$
real. This introduces non-Hermiticity into the system and allows us to
investigate whether the transition from integrability to chaos, previously
observed in the Hermitian limit, persists in the complex parameter space.

\begin{figure}
	\centering
	\includegraphics[width=0.9\linewidth]{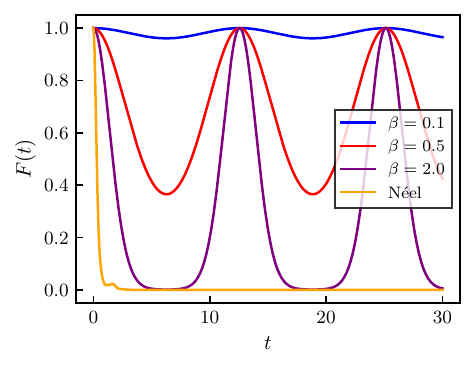}
	\caption{Fidelity dynamics of the Hamiltonian $H$ with parameters $(J_{h},J_{c}%
		,J_{z},J_{n},n)=(1,1,0.5,0.2i,3)$ under periodic boundary conditions, for a
		system size of $N=8$ and Hilbert-space dimension is $6561$. When the initial
		state is chosen as the coherent state defined in Eq.~(\ref{psibeta}), clear
		periodic revivals are observed, indicating that this coherent state, or
		equivalently the tower of states defined in Eq.~(\ref{eq:tower}), exhibits the
		characteristic behavior of a quantum many-body scar. In contrast, for the
		N\'{e}el state, the fidelity decays rapidly to zero, signaling fast
		thermalization and the absence of scar dynamics.}
	\label{fig:nh1_fidelity}
\end{figure}
We first analyze the non-Hermitian model with $%
(J_{h},J_{c},J_{z},n)=(1,1,0.5,3)$ and $J_{n}=0.2i$. The results, summarized
in Fig.~\ref{fig:nh1_krylov_energy}, demonstrate that the introduction of an
imaginary $J_{n}$ drives the system from integrability to chaos. The left
panel shows the time evolution of the Krylov complexity computed via Eq.~(%
\ref{krylov_complexity}), starting from an initial state given by an equal
superposition of all eigenstates in the $(M,k)=(0,0)$ subspace. The
complexity first rises, then decays, and finally saturates, consistent with
the dynamical signature of chaos. The right panel presents the CSR
distribution in the complex plane, which displays the expected Ginibre-type
repulsion pattern.

Having established the onset of chaos, we next explore whether the quantum
many-body scar (QMBS) states identified in the Hermitian limit persist in
this non-Hermitian setting. To this end, we compute the von Neumann
entanglement entropy using Eq.~(\ref{ent_entropy}) for the right
eigenvectors of the non-Hermitian Hamiltonian. As shown in Fig.~\ref%
{fig:nh1_entropy}, even across the integrable-to-chaotic transition, a
low-entropy state remains present. The coincidence between the low-entropy
states in both the Hermitian ($J_{n}=0$) and non-Hermitian ($J_{n}=0.2i$)
cases confirms the robustness of the scar tower described by Eq.~(\ref%
{eq:tower}).

Finally, to probe the dynamical properties of the scar states, we evaluate
the fidelity dynamics of the coherent state defined in Eq.~(\ref{psibeta})
and compare it with the N\'{e}el state. As shown in Fig.~\ref%
{fig:nh1_fidelity}, the coherent state exhibits clear periodic revivals, while the N\'{e}el
state rapidly thermalizes. These results confirm that the tower states $%
\{|\psi _{n}(N,N-p)\rangle \}$ continue to form a scar subspace even in the
non-Hermitian chaotic regime.

In summary, \textit{Case I} demonstrates that introducing a complex-valued
nonlocal coupling $J_n$ can drive the system from a Hermitian integrable
phase to a non-Hermitian chaotic regime, as verified by both the CRS and
Krylov-complexity diagnostics. To further reveal the universality of this
mechanism, in the next subsection we extend our analysis to \textit{Case II}%
, where the background Hamiltonian itself is already non-Hermitian (with
complex $J_c$), and examine whether the same $J_n$-induced perturbation can
again trigger the loss of integrability.

\subsection{\textit{Case II}: complex-valued $J_{c}$ and $J_{n}$}

In this subsection, we extend the analysis by allowing both $J_{c}$ and $%
J_{n}$ to take complex values. This generalization introduces an additional
source of non-Hermiticity, thereby enriching the dynamical behavior of the
model. As before, in order to identify potential quantum many-body scar
states, we first need to confirm that the system exhibits non-integrable
dynamics under certain parameter regimes.

\begin{figure}
\centering
\includegraphics[width=0.9\linewidth]{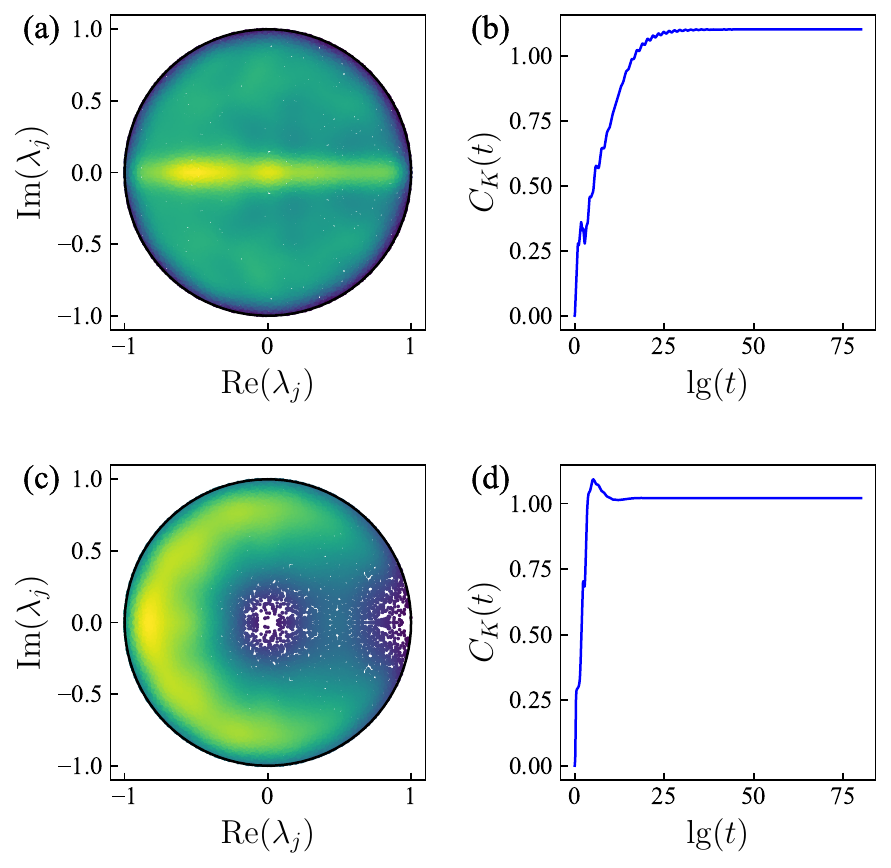}
\caption{Spectral and dynamical diagnostics of the non-Hermitian Hamiltonian 
$\mathcal{H}_{M=0,k=0}$ with parameters $(J_{h},J_{c},n)=(1,i,3)$. 
Panels~(a) and (b) display the CSR distribution and the time evolution of the 
Krylov complexity for $J_{n}=0$, respectively. The CSR shows an approximately 
uniform distribution in the complex plane, with 
$\langle\cos\theta_{j}\rangle \approx -0.028$, and the Krylov complexity grows 
linearly before saturating, both consistent with integrable behavior. 
Panels~(c) and (d) present the corresponding results for $J_{n}=0.2i$. In this 
case, the spectral distribution becomes strongly nonuniform, with 
$\langle\cos\theta_j\rangle \approx -0.187$, and the Krylov complexity exhibits 
rapid growth followed by decay and eventual saturation, indicating the onset of 
chaotic dynamics. Panels~(a) and (c) use a system size of $N=14$ (Hilbert-space 
dimension $44046$), while panels~(b) and (d) use $N=10$ (Hilbert-space dimension 
$902$).}
\label{fig:nh2_krylov_energy}
\end{figure}

To this end, we employ two complementary diagnostics: the level-spacing ratio
defined in Eq.~(\ref{CSR}) and the Krylov complexity defined in
Eq.~(\ref{krylov_complexity}). The results are summarized in
Fig.~\ref{fig:nh2_krylov_energy}. Specifically, panels (a) and (b) correspond
to the case where $J_{n}=0$, while panels (c) and (d) show results for
$J_{n}=0.2i$. The parameters are fixed as $(J_{h},J_{c},n)=(1,i,3)$, and all
calculations are performed within the $(M,k)=(0,0)$ subspace of the
Hamiltonian $\mathcal{H}_{M=0,k=0}$.

As $J_{n}$ increases from zero to a finite imaginary value, the system
undergoes a clear transition from integrability to chaos, similar to the
Hermitian case. Having established the chaotic nature of the model in this
parameter regime, we next explore the existence of quantum many-body scars.

\begin{figure*}[htbp]
\centering
\includegraphics[width=0.8\linewidth]{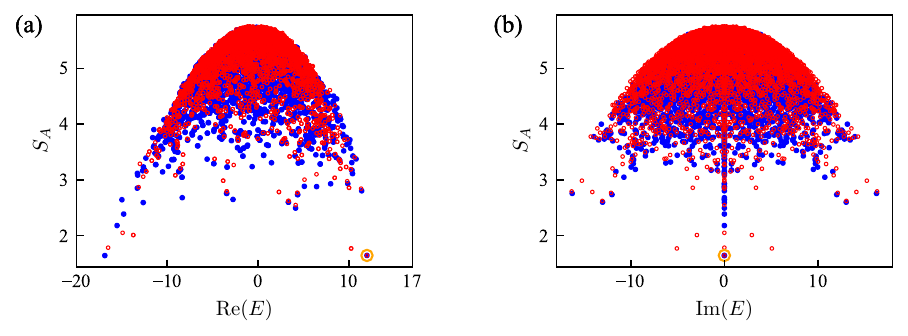}
\caption{
Von Neumann entanglement entropy computed via Eq.~(\ref{ent_entropy}) for the
right eigenvectors of $\mathcal{H}_{M=0,k=0}$ with parameters $(J_{h}%
,J_{c},J_{z})=(1,i,0.5)$ and $n=3$. The system size is $N=12$, with a
Hilbert-space dimension of $6166$. Panels~(a) and (b) show the entanglement
entropy plotted against the real and imaginary parts of the eigenvalues,
respectively. Blue dots correspond to the Hermitian case ($J_{n}=0$), while
red circles represent the non-Hermitian case ($J_{n}=0.2i$). In both parameter
regimes, a common low-entanglement eigenstate appears (highlighted in orange),
demonstrating the persistence and robustness of the quantum many-body scar
under non-Hermitian perturbations.
}

\label{fig:nh2_entropy}
\end{figure*}

To further characterize the structure of the eigenstates, we compute the von
Neumann entanglement entropy according to Eq.~(\ref{ent_entropy}), focusing
on the right eigenvectors of $\mathcal{H}_{M=0,k=0}$. The results, displayed
in Fig.~\ref{fig:nh2_entropy}, correspond to parameters $%
(J_{h},J_{c},J_{z},n)=(1,i,0.5,3)$ for a system size of $N=12$, within a
Hilbert space of dimension $6166$.

\begin{figure}[ptb]
\centering
\includegraphics[width=0.9\linewidth]{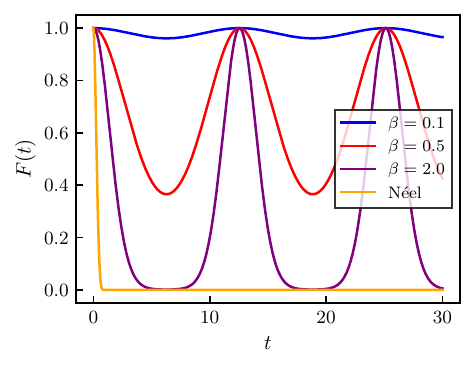}
\caption{Fidelity dynamics of the Hamiltonian $H$ with parameters $(J_{h},J_{c}%
	,J_{z},J_{n},n)=(1,i,0.5,0.2i,3)$ under periodic boundary conditions, for a
		system size of $N=8$ and Hilbert-space dimension is $6561$. When the initial
	state is chosen as the coherent state defined in Eq.~(\ref{psibeta}), clear
	periodic revivals are observed, indicating that this coherent state, or
	equivalently the tower of states defined in Eq.~(\ref{eq:tower}), exhibits the
	characteristic behavior of a quantum many-body scar. In contrast, for the
	N\'{e}el state, the fidelity decays rapidly to zero, signaling fast
	thermalization and the absence of scar dynamics.}
\label{fig:nh2_fidelity}
\end{figure}

The presence of low-entanglement eigenstates that remain robust as $J_{n}$
changes suggests the survival of a tower of scar states, consistent with the
analytical structure described by Eq.~(\ref{eq:tower}). To probe their
dynamical signatures, we evaluate the fidelity dynamics for both the
coherent scar state $|\psi_{\beta}\rangle$ defined in Eq.~(\ref{psibeta})
and the N\'{e}el state, as shown in Fig.~\ref{fig:nh2_fidelity}.

These results collectively show that the tower states $\{|\psi_{n}(N,N-p)%
\rangle\}$ retain their scar-like behavior even in the presence of
non-Hermiticity. The persistence of coherent revivals in fidelity and the
appearance of low-entanglement states confirm that the underlying scar
subspace remains robust against complex deformations of the coupling
parameters.

The two cases considered above together establish a coherent picture of the
non-Hermitian extension of our model. In \textit{Case I}, the system evolves
from a Hermitian integrable regime to a non-Hermitian chaotic one as the
complex hopping $J_n$ is introduced. In \textit{Case II}, the starting point
is already non-Hermitian but integrable (with complex $J_c$), and the same
complex $J_n$ again induces the transition to chaos. Despite the distinct
backgrounds, both cases reveal that $J_n$ acts as a universal source of
nonintegrability, controlling the crossover from regular to chaotic
dynamics. Remarkably, in both transitions the tower states survive as robust
quantum many-body scars, indicating that the scar subspace is protected not
only against Hermitian perturbations but also under generic non-Hermitian
deformations.

\section{Summary and Outlook}

\label{sec:summary}

In this work, we have systematically explored how the introduction of a
long-range nonlocal hopping term $H_{n}$ reshapes the integrability and
dynamical behavior of both Hermitian and non-Hermitian spin chains. In the
Hermitian regime, increasing $J_{n}$ first drives the system from an
integrable phase characterized by Poissonian level statistics to a chaotic
phase described by the Gaussian orthogonal ensemble, and eventually restores
integrability at large $J_{n}$. Extending the model into the non-Hermitian
domain reveals a unified mechanism: in \textit{Case I}, a complex $J_{n}$
transforms a Hermitian integrable system into a non-Hermitian chaotic one,
whereas in \textit{Case II}, an initially non-Hermitian but integrable
system loses integrability once the complex long-range coupling is
introduced. In both cases, $J_{n}$ acts as a universal control parameter
governing the transition from integrability to chaos.

Remarkably, despite these integrability-breaking transitions, a family of
QMBS states persists throughout both Hermitian and non-Hermitian regimes, maintaining coherent oscillatory dynamics and
revealing an unexpected robustness of nonthermal subspaces. This coexistence
of chaos and scars highlights the intricate interplay between nonlocality,
non-Hermiticity, and emergent order. Our results deepen the understanding of
integrability breaking in complex quantum systems and pave a new way to
controlling the quantum state in many-body systems.

\section*{DATA AVAILABILITY}

The data that support the findings of this article are openly available \cite{data}.

\acknowledgments 

We are grateful for the numerical calculation library Quspin \cite{quspin1,quspin2} which has facilitated our research. We also acknowledge the support of the National Natural Science Foundation of China under Grants No. 12275193 and 11975166.

\appendix

\section{Proof that the Tower States are Eigenstates}

\label{Proof that the Tower States are Eigenstates}

In this appendix, we prove that the tower of ferromagnetic states, defined
in the main text as Eq. (\ref{eq:tower}) are exact eigenstates of the full
Hamiltonian $H$. As shown in the main text, these states are trivially
eigenstates of the SU(2)-symmetric Heisenberg term $H_{h}$, the chiral
interaction $H_{c}$, and the Zeeman field $H_{z}$. The remaining task is to
demonstrate that they are also eigenstates of the nonlocal hopping term $%
H_{n}$. For a fixed hopping distance $n$ ($n\neq 0,N$ under PBCs), we define an operator (note that $H_{n}$ is a Hermitian operator, while $A_{n}$ is not).
\begin{equation}
A_{n}\equiv -iH_{n}=\sum_{j=1}^{N}\left(
S_{j}^{+}S_{j+n}^{-}-S_{j}^{-}S_{j+n}^{+}\right) .
\end{equation}%
If $\left\vert \psi \left( N,N-p\right) \right\rangle $ is an eigenstate of $%
A_{n}$, it is automatically an eigenstate of $H_{n}$. We now prove by
induction that 
\begin{equation}
A_{n}\left\vert \psi \left( N,N-p\right) \right\rangle =0,\qquad \forall
~p\geq 0.
\end{equation}%
For the fully polarized state $\left\vert \psi \left( N,N\right)
\right\rangle =\left\vert \Uparrow \right\rangle $, 
\begin{equation}
A_{n}\left\vert \Uparrow \right\rangle =\sum_{j=1}^{N}\left(
S_{j}^{+}S_{j+n}^{-}-S_{j}^{-}S_{j+n}^{+}\right) \left\vert \Uparrow
\right\rangle =0,
\end{equation}%
since $S_{j}^{+}\left\vert +1\right\rangle _{j}=0$ and $S_{j+n}^{+}\left%
\vert +1\right\rangle _{j+n}=0$. Hence, $\left\vert \Uparrow \right\rangle $
is annihilated by $A_{n}$.

The commutator between $A_{n}$ and the total lowering operator $%
S^{-}=\sum_{m}S_{m}^{-}$ reads 
\begin{equation}
\left[ A_{n},S^{-}\right] =\sum_{j=1}^{N}\left[ \left(
S_{j}^{+}S_{j+n}^{-}-S_{j}^{-}S_{j+n}^{+}\right) ,S_{j}^{-}+S_{j+n}^{-} %
\right] ,
\end{equation}
where operators on different sites commute. Using $[S_{j}^{+},S_{k}^{-}]=2
\delta _{jk}S_{j}^{z}$ and $[S_{j}^{-},S_{k}^{-}]=0$, we obtain 
\begin{equation}
\left[ A_{n},S^{-}\right] =2\sum_{j=1}^{N}\left(
S_{j}^{z}S_{j+n}^{-}-S_{j}^{-}S_{j+n}^{z}\right) .  \label{eq:comm1_app}
\end{equation}
Acting this on $\left\vert \Uparrow \right\rangle $ gives 
\begin{align}
\left[ A_{n},S^{-}\right] \left\vert \Uparrow \right\rangle &
=2\sum_{j=1}^{N}\left( S_{j+n}^{-}-S_{j}^{-}\right) \left\vert \Uparrow
\right\rangle  \notag \\
& =2\left( \sum_{j=1}^{N}S_{j+n}^{-}-\sum_{j=1}^{N}S_{j}^{-}\right)
\left\vert \Uparrow \right\rangle .
\end{align}
Under periodic boundary conditions, $\sum_{j}S_{j+n}^{-}=\sum_{j}S_{j}^{-}$,
and therefore 
\begin{equation}
\left[ A_{n},S^{-}\right] \left\vert \Uparrow \right\rangle =0.
\label{eq:commpsi_app}
\end{equation}

Taking a further commutator yields 
\begin{equation}
\left[ \left[ A_{n},S^{-}\right] ,S^{-}\right] =2\sum_{j=1}^{N}\left[
S_{j}^{z}S_{j+n}^{-}-S_{j}^{-}S_{j+n}^{z},S_{j}^{-}+S_{j+n}^{-}\right] .
\end{equation}
Each term in the summation cancels exactly, giving the operator identity 
\begin{equation}
\left[ \left[ A_{n},S^{-}\right] ,S^{-}\right] =0.  \label{eq:comm2_app}
\end{equation}

Equations~(\ref{eq:comm1_app}) and~(\ref{eq:comm2_app}) imply that $\left[
A_{n},S^{-}\right] \left\vert \Uparrow \right\rangle =0$ and $%
[\,[A_{n},S^{-}],S^{-}]=0$. Let $X=[A_{n},S^{-}]$; then $[X,S^{-}]=0$, which
means $[X,(S^{-})^{k}]=0$ for any $k\geq 1$. Applying this to $%
(S^{-})^{p-1}\left\vert \Uparrow \right\rangle $ yields 
\begin{equation}
\left[ A_{n},S^{-}\right] (S^{-})^{p-1}\left\vert \Uparrow \right\rangle =0.
\end{equation}
Expanding the commutator gives 
\begin{equation}
A_{n}(S^{-})^{p}\left\vert \Uparrow \right\rangle
-S^{-}A_{n}(S^{-})^{p-1}\left\vert \Uparrow \right\rangle =0.
\end{equation}
It follows recursively that 
\begin{equation}
A_{n}\left\vert \psi \left( N,N-p\right) \right\rangle =0,\quad \forall
~p\geq 0.
\end{equation}
Since $H_{n}=iA_{n}$, one has 
\begin{equation}
H_{n}\left\vert \psi \left( N,N-p\right) \right\rangle =iA_{n}\left\vert
\psi \left( N,N-p\right) \right\rangle =0.
\end{equation}
Therefore, the tower states $\left\vert \psi \left( N,N-p\right)
\right\rangle $ are eigenstates of $H_{n}$ with eigenvalue zero. Combined
with their trivial eigenstate property under $H_{h}$, $H_{c}$, and $H_{z}$,
this completes the proof that all $\left\vert \psi \left( N,N-p\right)
\right\rangle $ are exact eigenstates of the full Hamiltonian $H$.

\section{Bi-Lanczos Algorithm for Non-Hermitian Systems}

\label{Bi-Lanczos Algorithm for Non-Hermitian Systems}

In order to analyze the Krylov complexity in non-Hermitian quantum systems,
we employ the bi-Lanczos algorithm, which generalizes the conventional
Lanczos recursion to handle non-Hermitian Hamiltonians. The method
constructs two biorthogonal Krylov bases, $\{ |p_n\rangle \}$ and $\{
|q_n\rangle \}$, corresponding to the right and left subspaces of the
Liouvillian or Hamiltonian operator.

\subsection{Algorithmic Construction}

\label{Algorithmic Construction}

We start from an arbitrary normalized initial state $|\psi_0\rangle$, and
define the initial vectors of the right and left Krylov subspaces as 
\begin{equation}
|p_0\rangle = |\psi_0\rangle, \quad |q_0\rangle = |\psi_0\rangle,
\end{equation}
with the normalization condition $\langle q_0 | p_0 \rangle = 1$.

At each recursion step $n$, we apply both the Hamiltonian $H$ and its
Hermitian conjugate $H^\dagger$ to the basis vectors: 
\begin{equation}
|p_{n+1}\rangle = H |p_n\rangle, \qquad |q_{n+1}\rangle = H^\dagger
|q_n\rangle.
\end{equation}

To enforce biorthogonality, we define the Lanczos coefficients as 
\begin{equation}
a_{n}=\langle q_{n}|p_{n+1}\rangle ,
\end{equation}%
and remove the components along the previous two Krylov vectors, 
\begin{align}
|p_{n+1}\rangle & \leftarrow |p_{n+1}\rangle -a_{n}|p_{n}\rangle
-c_{n}|p_{n-1}\rangle , \\
|q_{n+1}\rangle & \leftarrow |q_{n+1}\rangle -a_{n}^{\ast }|q_{n}\rangle
-b_{n}^{\ast }|q_{n-1}\rangle ,
\end{align}%
where $b_{n}$ and $c_{n}$ are the normalization coefficients ensuring that $%
\langle q_{n+1}|p_{n+1}\rangle =1$. Specifically, 
\begin{equation}
b_{n}=\Vert |p_{n+1}\rangle \Vert ,\qquad c_{n}=\langle
q_{n+1}|p_{n+1}\rangle /b_{n}.
\end{equation}

The vectors are then normalized as 
\begin{equation}
|p_{n+1}\rangle \leftarrow \frac{|p_{n+1}\rangle}{b_n}, \qquad
|q_{n+1}\rangle \leftarrow \frac{|q_{n+1}\rangle}{c_n^*}.
\end{equation}

In practice, due to the non-unitary evolution in non-Hermitian systems,
small numerical errors can accumulate, breaking the biorthogonality
condition. To mitigate this, a reorthogonalization procedure is performed
after each iteration, ensuring 
\begin{equation}
\langle q_i | p_j \rangle = \delta_{ij}.
\end{equation}

\subsection{Effective Tridiagonal Representation}

The bi-Lanczos recursion leads to a tridiagonal effective representation of
the Hamiltonian in the biorthogonal Krylov basis: 
\begin{equation}
H|p_{n}\rangle =b_{n+1}|p_{n+1}\rangle +a_{n}|p_{n}\rangle
+c_{n}|p_{n-1}\rangle ,
\end{equation}%
and similarly for the adjoint relation 
\begin{equation}
H^{\dagger }|q_{n}\rangle =c_{n+1}^{\ast }|q_{n+1}\rangle +a_{n}^{\ast
}|q_{n}\rangle +b_{n+1}^{\ast }|q_{n-1}\rangle .
\end{equation}%
Thus, the projected Hamiltonian in the Krylov subspace takes the matrix form 
\begin{equation}
H_{\mathrm{eff}}=%
\begin{pmatrix}
a_{0} & c_{1} & 0 & \cdots & 0 \\ 
b_{1} & a_{1} & c_{2} & \cdots & 0 \\ 
0 & b_{2} & a_{2} & \cdots & 0 \\ 
\vdots & \vdots & \vdots & \ddots & c_{d-1} \\ 
0 & 0 & 0 & b_{d-1} & a_{d-1}%
\end{pmatrix}%
.
\end{equation}

\subsection{Connection to Krylov Complexity}

Once the biorthogonal Krylov bases $\{|p_{n}\rangle ,|q_{n}\rangle \}$ are
constructed, the time-evolved state can be expressed as 
\begin{equation}
|\psi (t)\rangle =e^{-iHt}|\psi _{0}\rangle =\sum_{n}\phi
_{n}(t)|p_{n}\rangle ,
\end{equation}
where the coefficients $\phi _{n}(t)$ obey a set of coupled differential
equations governed by $H_{\mathrm{eff}}$. The Krylov complexity, which
measures the spread of the state in the Krylov basis, is then given by 
\begin{equation}
C_{K}(t)=\sum_{n}n\,|\phi _{n}(t)|^{2}.
\end{equation}
For non-Hermitian systems, both left and right Krylov amplitudes contribute,
and the norm is generalized to $\langle \psi (t)|\psi (t)\rangle
=\sum_{n}|\phi _{n}(t)|^{2}$ in the biorthogonal sense.

\end{document}